\DeclareUnicodeCharacter{2062}{} % Fix for INVISIBLE TIMES
\documentclass[journal=ancac3, manuscript=article]{achemso}

\usepackage{graphicx}% Include figure files
\usepackage{xcolor}
\usepackage{bm}% bold math
\usepackage{siunitx}
\DeclareSIUnit\angstrom{\text{\AA}}
\usepackage{hyperref}
\usepackage{xr-hyper}
\usepackage{ulem}
\usepackage{amsmath}
\usepackage{booktabs}
\usepackage{subcaption}
\definecolor{rev_1}{rgb}{0.00,0.00,1.00} 
\title{Composition-controlled optical fingerprints and anisotropy of self-trapped excitons in double perovskites
}
\author{Simone Grillo}
\affiliation{Max Planck Institute for the Structure and Dynamics of Matter (MPSD) and Center for Free-Electron Laser Science (CFEL), 22761 Hamburg, Germany}
\alsoaffiliation{These authors contributed equally}
\email{simone.grillo@mpsd.mpg.de}
\author{Offir Zachs Maishlos}
\affiliation{Department of Materials Science and Engineering
Technion - Israel Institute of Technology, Haifa 32000, Israel}
\alsoaffiliation{These authors contributed equally}
\author{Liya Koren}
\affiliation{Department of Materials Science and Engineering
Technion - Israel Institute of Technology, Haifa 32000, Israel}
\author{Rotem Strassberg}
\affiliation{Department of Materials Science and Engineering
Technion - Israel Institute of Technology, Haifa 32000, Israel}
\alsoaffiliation{The Solid-State Institute Technion - Israel Institute of Technology, Haifa 32000, Israel}
\author{Rachel Lifer}
\affiliation{Department of Materials Science and Engineering
Technion - Israel Institute of Technology, Haifa 32000, Israel}
\author{Edi Maizenberg}
\affiliation{Department of Materials Science and Engineering
Technion - Israel Institute of Technology, Haifa 32000, Israel}
\author{Umberto De Giovannini}
\affiliation{Dipartimento di Fisica e Chimica - Emilio Segrè, Università degli Studi di Palermo, Palermo I-90123, Italy}
\author{Angel Rubio}
\affiliation{Max Planck Institute for the Structure and Dynamics of Matter (MPSD) and Center for Free-Electron Laser Science (CFEL), 22761 Hamburg, Germany}	
\alsoaffiliation{Nano-Bio Spectroscopy Group and ETSF, Universidad del País Vasco UPV/EHU, 20018 San Sebastián, Spain}
\alsoaffiliation{Initiative for Computational Catalysis (ICC) and Center for Computational Quantum Physics (CCQ), The Flatiron Institute, Simons Foundation, New York City, NY 10010, United States of America}
\email{angel.rubio@mpsd.mpg.de}
\author{Yehonadav Bekenstein}
\affiliation{Department of Materials Science and Engineering
Technion - Israel Institute of Technology, Haifa 32000, Israel}
\alsoaffiliation{The Solid-State Institute Technion - Israel Institute of Technology, Haifa 32000, Israel}
\email{bekenstein@technion.ac.il}
\author{Ofer Neufeld}
\affiliation{Technion - Israel Institute of Technology, Schulich Faculty of Chemistry, Haifa 3200003, Israel}
\email{ofern@technion.ac.il}

\begin{document}
% \maketitle
\singlespace

\vspace{-1cm}
\begin{abstract}

Double perovskites (DPs) are an emerging functional materials platform for photocatalysis and optoelectronics, offering greater compositional flexibility than their single-perovskite counterparts. Like single perovskites, DPs possess soft lattices that promote strong exciton–phonon coupling and formation of self-trapped excitons (STEs), giving rise to broad photoluminescence (PL) at energies below the absorption onset. Here we experimentally investigate lead-free Sb-doped Cs$_2$BInCl$_6$ DP nanocrystals, in which the B site is alloyed from Na to K. Optical measurements reveal tunable PL from $\sim$2.48 to $\sim$2.70 eV upon alloying, consistent with a homogeneous solid solution. In contrast, absorption features near $\sim$3.8 eV remain largely insensitive to composition.
We investigate the microscopic origin of this behavior using \textit{ab initio} theory. Our calculations show that Sb doping introduces high oscillator strength radiative recombination channels and delocalizes the excitonic wavefunction. We further identify STEs that substantially red-shift PL, driven by an upward shift of dopant-hybridized in-gap states. Substitution from Na to K on the B site selectively modifies the STE lattice distortion, thereby tuning the PL while leaving absorption comparatively unchanged. Lastly, our \textit{ab initio} theory predicts extremely strong PL polarization anisotropies that directly reflect the spatial anisotropy of the STE, depending on the B-site composition. Time-resolved anisotropy measurements perfectly match the theory, providing first experimental fingerprints of the excitonic wavefunction in DP and its composition dependence. Our results provide \textit{ab initio} insights into STEs in lead-free DPs and clarify the distinct roles of dopants and alloying in controlling tunable PL. These findings offer a framework for interpreting experiments and pave the way to novel optical spectroscopies for reconstructing the excitonic wavefunction structure.

\end{abstract}
\newpage

\section{Introduction}
Over the past decade, double perovskites (DPs) have been proposed as candidate materials \cite{Slavney2016} and nanomaterials \cite{Bekenstein2018} for several emerging technologies, from photovoltaic applications \cite{Nair2022} to optoelectronics \cite{Chu2019} and spintronics \cite{kepenekian2017rashba}. The main advantage of DPs over single perovskites (SPs) is the additional structural degree of freedom obtained by doubling the unit cell (from ABX$_3$ to A$_{2}$BB$^\prime$X$_{6}$, with A and B indicating the cationic sites and X the halides), allowing arrangements of different cations \cite{Anderson1993, Muscarella2022}. This provides potential tunability in various material properties such as gaps, stability, rigidity, phononic bands, etc., which is appealing for applications. In this context, lead-free systems are especially promising, as they are non-toxic and provide stronger commercial viability \cite{gray2020high}. 

Excitonic properties of perovskites have attracted significant attention from the scientific community in recent years, both in theory \cite{filip2021phonon, quarti2024exciton} and experiments \cite{baranowski2020excitons, baek2025exciton}. Indeed, excitons typically dominate absorption and luminescence spectra, directly impacting optoelectronics. One of the most striking features is the so-called self-trapped excitons (STEs) \cite{yu2026perspective}, which somewhat resemble small-polarons but with an electron-hole wavefunction coupling to the lattice instead of a band electron. Here, following excitation (promotion of an electron to the conduction band via photon absorption, and subsequent excitonic binding to the valence hole) \cite{Gill2025,Chen2026}, the crystal lattice deforms around the exciton \cite{yu2026perspective}. This process reduces the total energy of the excited electron-nuclei system and is enabled by strong exciton-phonon coupling in the soft lattice \cite{chen2023strong}. STEs are known to be highly localized in SPs, and their telltale signature is a red-shifting of the photoluminescence (PL) peak compared to the optical absorption threshold \cite{luo2018efficient}, indicating that the lattice undergoes a structural change in between the two processes \cite{yu2026perspective}. This can be accompanied by lattice symmetry breaking and anisotropic emission polarization \cite{Wang2022}. On the other hand, much less is known about self-trapped states in DP. These have been discussed experimentally and interpreted in terms of self-trapping based on the pronounced red-shifting effect, supported by theoretical analyses \cite{Noculak2020}. However, first-principles descriptions of STEs in DPs at the level of many-body excited-state theory are lacking \cite{wang2019atomistic, jin2024self, yu2026perspective}. Main questions of interest include what is the level of localization of STEs in DP, how do they depend on the cation choices and ratios, and what role do dopants play \cite{Khalfin2019, Noculak2020, chen2023strong}. It is also unknown how composition degrees of freedom map onto the STE wavefunction and can be optically probed. Addressing these questions is ultimately essential for developing improved materials platforms. 

Here we perform a combined experimental and theoretical investigation of excitons in DP. We synthesize Sb-doped Cs$_2$BInCl${_6}$ nanoparticles alloyed from B$\,=\,$Na to B$\,=\,$K through all compositions (going beyond past efforts \cite{Dengo2025}). We measure absorption, photoluminescence excitation (PLE), and PL, and demonstrate that alloying continuously tunes PL from $\sim$2.48 to $\sim$2.7 eV, while absorption peaks appear at much higher energy and their position is independent of composition. Our measurements show that strong PL is only obtained in doped samples, suggesting that the dopant creates an effective excitonic recombination site in the distorted lattice. To understand the fundamental physical mechanism behind these features, we perform state-of-the-art \textit{ab initio} calculations that involve many-body perturbation theory (MBPT), including $G_0W_0$ quasiparticle corrections and the Bethe-Salpeter equation (BSE), and constrained spin density functional theory (cDFT). Our simulations confirm that the self-trapping mechanism in DPs is indeed responsible for the red-shifted PL. Excitons are found to be highly localized in the pristine crystal, while Sb doping causes delocalization due to hybridization with flat in-gap defect-like states. We uncover that lattice-exciton coupling is selective of the DP cation composition, with B$\,=\,$K inducing a stronger distortion of the lattice that enables more pronounced red-shifted PL due to relative motion of dopant, Cl, and In, hybridized in-gap states. Lastly, we predict from \textit{ab initio} simulations, and validate experimentally, that STE PL in DP is highly polarization anisotropic, with the anisotropy directly reflecting the spatial structure of the STE wavefunction. Alloying the B site provides control over the anisotropy and allows probing the excitonic state structure. Our work provides valuable insight into DP excitonic physics, and proposes a novel scheme for optical probing of excitonic wavefunctions. 

\section{Results and Discussion}

\subsection{Linear-response optical measurements}

Let us begin by describing the experimental set-up. We use the solution-based hot injection method in inert conditions to synthesize stable colloidal nanoparticles of different compositions of Cs$_2$Na$_x$K$_{1-x}$In$_{0.9}$Sb$_{0.1}$Cl$_6$. The synthesis follows the approach of Locardi et al. \cite{locardi2019emissive}. All additional technical details about sample preparation and the experimental procedures are delegated to the Materials and Methods section.  

\begin{figure*}
    \centering
    \includegraphics[scale=0.486]{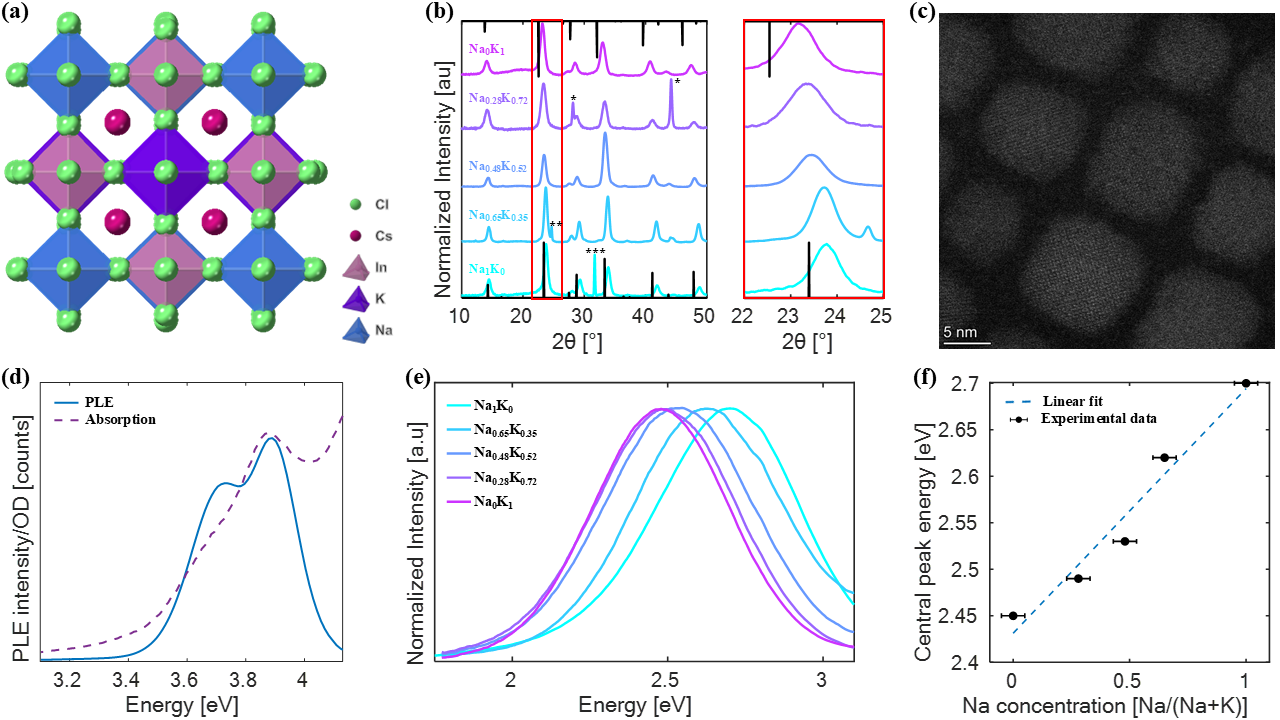}
    \caption{[\textbf{a}] Schematic structure of alloyed Cs$_2$Na$_x$K$_{1-x}$In$_{0.9}$Sb$_{0.1}$Cl$_6$ for $x=0.5$ \cite{CrystalMaker, MaterialsProject_mp989571}. [\textbf{b}] X-ray diffraction patterns of Cs$_2$Na$_x$K$_{1-x}$In$_{0.9}$Sb$_{0.1}$Cl$_6$ nanocrystals, with $x=0,0.28,0.48,0.63,1$. The black traces show the calculated XRD patterns of Cs$_2$KInCl$6$ (mp-1112759, top) and of Cs$_2$NaInCl$6$ (mp-989571, bottom) from the Materials Project website. Additional sharp peaks, assigned to different salt byproducts of the synthesis, are marked with * (KCl), ** (CsCl), and *** (NaCl). The right panel shows a zoomed-in view of the $2\theta=22^\circ-25^\circ$ range,highlighting the linear peak shift with composition. [\textbf{c}] HR-TEM micrograph of Cs$_2$KIn$_{0.9}$Sb$_{0.1}$Cl$6$ nanocrystals. [\textbf{d}] Photoluminescence-excitation (PLE) spectrum (solid blue line) and absorption spectrum profile (dotted purple line), almost independent of the nanocrystal composition. [\textbf{e}] PL emission spectra of Cs$_2$Na$_x$K$_{1-x}$In$_{0.9}$Sb$_{0.1}$Cl$_6$ nanocrystals, with $x=0,0.28,0.48,0.63,1$, measured under \SI{320}{\nano\meter} excitation. [\textbf{f}] PL central peak energy as a function of the Na$^+$ fraction. The dotted blue line is a linear fit $(R^2=0.96)$.
    \label{fig_1}}
\end{figure*}

Fig.~\ref{fig_1} shows our main experimental results. High-resolution transmission electron microscope (HR-TEM) micrographs present cubic-shaped Cs$_2$KIn$_{0.9}$Sb$_{0.1}$Cl$_6$ nanoparticles of 10-15 nm in size (Fig.~\ref{fig_1}c; micrographs for more samples are shown in Fig.~S9 of the Supporting Information, SI). We chose to use a constant 10\% Sb doping throughout experiments since in our previous studies this ratio was found optimal for optical properties of the Cs$_2$NaIn$_{1-x}$Sb$_x$Cl$_6$ system \cite{shaek2023does}. Note that in all cases, we do not observe significant PL in undoped samples, indicating that the dopant plays a critical role in either exciton formation, or recombination.   
We verify the Na/K ratios in samples using SEM-EDS (see Fig.~S8 in the SI), while X-ray diffraction (XRD) measurements were used to verify the crystal structure of samples. All compositions were found to crystallize in the cubic phase with \textit{Fm}$\bar{3}$\textit{m} symmetry. Subsequently, XRD peaks were analyzed using a pseudo-Voigt function. The extracted lattice parameters were found to exhibit a linear decrease with increasing Na concentration, consistent with Vegard’s law \cite{vegard1921konstitution}.

Next, Fig.~\ref{fig_1}\textit{d} presents absorption PLE measurements. In all compositions the main absorption peak is located at 3.87 eV (320 nm), with an additional peak at 3.65 eV (340 nm). Those two peaks match prior observations of dynamic Jahn-Teller distortion of the excited states \cite{Noculak2020, gray2020high}.
On the one hand, for Sb-doped samples within the full alloying range (from B$\,=\,$K to B$\,=\,$Na), we find that PLE measurements (which follow absorption spectra) are almost independent of the composition, with only the intensity ratio between the two peaks varying. 
This suggests that Na/K elements in the DP crystal do not directly affect the exciton generation process and light absorption \cite{zhu2021sb}, despite the slight variation in lattice parameter (from \SI{10.58}{\angstrom} to \SI{10.82}{\angstrom}, see Table~S1 in the SI).
On the other hand, PL measurements uncover a completely different story. Here, Fig~\ref{fig_1}e shows that the PL is red-shifted compared to absorption peaks, creating a Stokes shift up to 1.2 eV, which is further strongly dependent on composition.
This is often a telltale sign for STEs, together with a $\mu$s lifetime regime (see SI for further discussion). Notably, alloying between the B$\,=\,$K and B$\,=\,$Na samples with varying composition (Cs$_2$K$_x$Na$_{1-x}$InCl${_6}$) smoothly shifts the PL emission peak by as much as $\sim 0.2$ eV \cite{Noculak2020}. This indicates that the different cations strongly alter the electronic properties of the excitons upon emission, i.e. they might alter the structure of the self-trapped states that have been formed after photo-excitation.

\subsection{Ab initio STE calculations}

Next, we perform an \textit{ab initio} analysis of optoelectronic properties of DPs. We first investigate the absorption spectra of Cs$_2$NaInCl$_6$ and Cs$_2$KInCl$_6$ DPs, both pristine and Sb-doped, explicitly including excitonic effects. Owing to computational constraints, the Sb-doped systems are modeled using a dopant concentration of 25\%. Although higher than the experimental samples, previous work has shown that increasing the Sb concentration only slightly affects the lattice parameter, consistent with the similar ionic radii of Sb$^{3+}$ and In$^{3+}$, $r(\mathrm{Sb}^{3+}) = 76$ pm and $r(\mathrm{In}^{3+}) = 80$ pm. More importantly, the trends in optoelectronic properties remain essentially unchanged, with Sb concentration primarily affecting the PL intensity \cite{shaek2023does}.

The simulations are performed by first relaxing the crystal structures at the DFT level using the GGA-PBEsol exchange-correlation (XC) functional. The systems are modeled in cubic \textit{Fm}$\bar{3}$\textit{m} cells, while further technical details are provided in the Materials and Methods section. The relaxed lattice parameters are reported in Table~S1 of the SI and are in good agreement with previous theoretical and experimental studies \cite{Noculak2020, shakeel2024dft, zhou2019manipulation, khan2021elastic, gray2020high, jeevaraj2024extrinsic, rameshkumar2023exploring, guedira1988phase}. Sb doping does not alter the average crystal structure, but it induces a slight increase in the lattice parameter, as expected \cite{shannon1976revised, tran2017designing, gray2020high}.
We further verified that spin-orbit coupling and semicore states have negligible effects on the electronic structure, as shown in Fig.~S1 of the SI. These effects are therefore neglected in the following analysis.

\begin{figure*}
    \centering
    \includegraphics[scale=0.59]{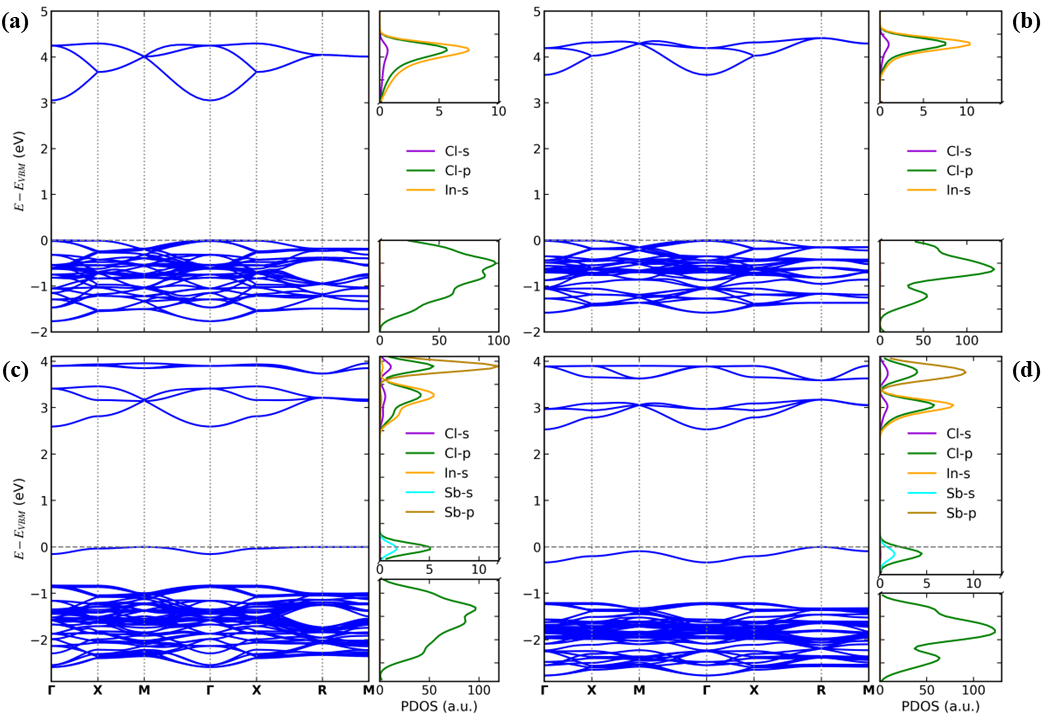}
    \caption{Electronic band structures and PDOS at the DFT level of [\textbf{a}] Cs$_2$NaInCl${_6}$, [\textbf{b}] Cs$_2$KInCl${_6}$, [\textbf{c}] Cs$_2$NaSb$_{0.25}$In$_{0.75}$Cl${_6}$, and [\textbf{d}] Cs$_2$KSb$_{0.25}$In$_{0.75}$Cl${_6}$. The zero reference is set at the VBM. 
    \label{fig_2}}
\end{figure*}

The electronic structures of the relaxed systems at the DFT level are presented in Fig.~\ref{fig_2}. The pristine compounds both exhibit a direct gap at $\Gamma$ (see Table~\ref{tab_1}), with flat band dispersion along both $\Gamma-$X and $\Gamma-$M high-symmetry directions \cite{zhou2019manipulation}. The valence bands are dominated by Cl-\textit{p} orbitals, while conduction bands primarily comprise hybridized In-\textit{s} and Cl-\textit{p} orbitals. Sb doping causes the emergence of a pronounced flat in-gap defect band, arising from hybridized Sb-\textit{s} and Cl-\textit{p} states (Fig.~\ref{fig_2}\textit{c}-\textit{d}), which significantly reduces the gap. We also note that both Na and K cation orbitals contribute negligibly to the valence and conduction edges. From this analysis, we can indeed expect a largely similar optical absorption in both cases, as well as similar excitonic wavefunctions following light absorption.

\begin{table}[h!]
    \centering
    \caption{Estimate of the lowest electronic gap of (Sb-doped) Na- and K-based DPs, using a GGA-PBEsol XC functional, at the DFT ($E^{\mathrm{DFT}}$) and $G_0W_0$ ($E^{G_0W_0}$) level. Also, the optical gap ($E_{o}^{\mathrm{BSE@}G_0W_0}$) and the binding energy ($E_{B}^{\mathrm{BSE@}G_0W_0}$) of the first bright exciton at the BSE@$G_0W_0$ level are displayed. For indirect electronic gaps, the lowest direct electronic gaps are shown in round brackets.}
    \label{tab_1}
    \begin{tabular}{lcccc}
        \toprule
        \textbf{System} & \textbf{E$^{\mathbf{DFT}}$} (eV) & \textbf{E$^{\mathbf{G_0W_0}}$} (eV) & \textbf{E$_\mathbf{o}^{\mathbf{BSE@G_0W_0}}$} (eV) & \textbf{E$^{\mathbf{BSE@G_0W_0}}_\mathbf{B}$} (eV) \\
        \midrule
        Cs$_2$NaInCl$_6$ & 3.05 & 5.62 & 5.26 & 0.36 \\
        Cs$_2$KInCl$_6$ & 3.61 & 6.32 & 5.68 & 0.64 \\
        Cs$_2$NaSb$_{0.25}$In$_{0.75}$Cl$_6$ & 2.59 (2.75) & 4.60 (4.73) & 4.34 & 0.39 \\
        Cs$_2$KSb$_{0.25}$In$_{0.75}$Cl$_6$ & 2.53 (2.87) & 4.69 (5.02) & 4.59 & 0.43 \\
        \bottomrule
    \end{tabular}
\end{table}

\begin{figure*}
    \centering
    \includegraphics[scale=0.52]{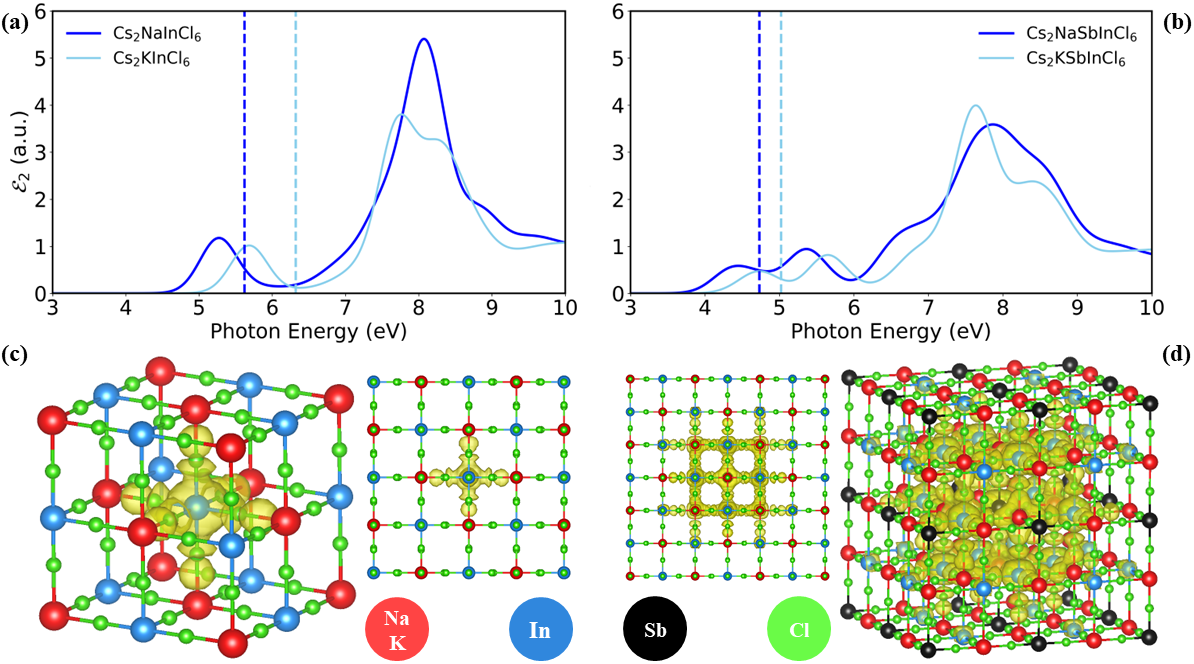}
    \caption{Optical absorption spectra, calculated at the BSE@$G_0W_0$ level as the imaginary part of the dielectric function, of [\textbf{a}] Cs$_2$NaInCl${_6}$ (blue) and Cs$_2$KInCl${_6}$ (light blue), and [\textbf{b}] Cs$_2$NaSb$_{0.25}$In$_{0.75}$Cl${_6}$ (blue) and Cs$_2$KSb$_{0.25}$In$_{0.75}$Cl${_6}$ (light blue). The relative $G_0W_0$ direct band gaps are also highlighted. A Gaussian broadening of 0.25 eV was used. Real-space representation of the wavefunctions of the first bright exciton of [\textbf{c}] Cs$_2$NaInCl${_6}$ and Cs$_2$KInCl${_6}$, and [\textbf{d}] Cs$_2$NaSb$_{0.25}$In$_{0.75}$Cl${_6}$ and Cs$_2$KSb$_{0.25}$In$_{0.75}$Cl${_6}$ are also shown. Cs atoms were omitted for clarity. 
    \label{fig_3}}
\end{figure*}

From the DFT starting point, we perform a single-shot $GW$ (i.e., $G_0W_0$) quasiparticle correction based on Green's function formalism \cite{hedin1965new} ($G_0W_0$ calculations are performed using Yambo code \cite{marini2009yambo,sangalli2019many}). This method corrects single-particle energies and band gaps of the Kohn-Sham (KS) eigenstates obtained from the DFT as a first step of the MBPT to study excited state properties. Our findings are in good agreement with previous $G_0W_0$ calculations for Cs$_2$NaInCl$_6$ \cite{leppert2024excitons, biega2023chemical}, validating the approach. The next stage of the MBPT framework is to account for electron-hole interactions by solving the BSE, using the $G_0W_0$-corrected states (BSE@$G_0W_0$) to calculate optical absorption spectra including excitonic features (see Fig.~\ref{fig_3}\textit{a}-\textit{b}, and Table~\ref{tab_1}). The pristine systems both exhibit a single relatively intense below-gap excitation at 5.26 eV for Cs$_2$NaInCl$_6$, and at 5.68 eV for Cs$_2$KInCl$_6$ with exciton binding energies of 0.36 eV and 0.64 eV, respectively. Sb-doping mostly induces a rigid red-shift of about 1 eV to the excitonic peaks compared to the pristine systems, that is, 4.34 eV for Cs$_2$NaSb$_{0.25}$In$_{0.75}$Cl$_6$ and 4.59 eV for Cs$_2$KSb$_{0.25}$In$_{0.75}$Cl$_6$, yielding binding energies of 0.39 eV and 0.43 eV, respectively. This means that, at the absorption level, doping does not play a major role. Simulations are overall in good agreement with experimental measurements, up to a constant shift of 0.5 eV. Most importantly, in both systems, the absorption spectra are roughly independent of the chemical composition (only 0.25 eV difference in the absorption edge between Na- and K-based DPs), as expected from PLE measurements (see Fig.~\ref{fig_1}), and from the fact that the DOS of Na and K cations is negligible at the band gap edges (see Fig.~\ref{fig_2}). Hence, these results give us confidence that the \textit{ab initio} simulations reliably capture the main physical effects in this system.

Further analyzing the spectrum, we find that the first bright excitations do not originate from direct transitions between the valence band maxima (VBM) and conduction band minima (CBM) since they are dipole-forbidden (i.e., optically dark) at $\Gamma$ in both compounds. This is a common feature in DPs \cite{meng2017parity, luo2018cs2agincl6, zhou2019manipulation}. Instead, in the pristine system the first bright exciton arises from transitions between deeper valence states and the lowest conduction bands along the $\Gamma$-X-R path in the BZ (see Fig.~S3\textit{a}-\textit{b} in the SI), and is a three-fold degenerate \textit{p}-like exciton \cite{grosso2013solid}. A real-space representation of the excitonic wavefunction in the pristine systems is shown in Fig.~\ref{fig_3}\textit{c}, where the \textit{p}-like nature of the exciton is clearly visible. In the Sb-doped compounds, the exciton retains a predominantly $p$-like character arising from transitions between valence states with hybridized Sb-$s$ and Cl-$p$ orbitals and conduction states with hybridized In-$s$ and Cl-$p$ orbitals, mainly along the $\Gamma$-X direction (see Fig.~S3\textit{c}-\textit{d} in the SI). Stemming from a very flat Sb-rich valence band in reciprocal space, the excitonic wavefunction is much more delocalized (extended) in real space in the doped case, as shown in Fig.~\ref{fig_3}\textit{d}. Thus, even though the absorption edges change negligibly with doping, the exciton characteristics greatly differ, including oscillator strengths and coupling to the lattice, which also affects PL spectra as we show below.

At this stage, we recall that absorption and PL can be regarded as time-reversed processes. Accordingly, the PL emission peak is expected to occur at energies close to the absorption edge. However, experiments consistently show a pronounced redshift of the PL spectra with respect to absorption (see Fig.~\ref{fig_1}).
This discrepancy suggests that PL does not originate from the same crystal configuration as the initial absorption process. Instead, following photoexcitation the system could undergo rapid structural relaxation, leading to a distortion of the lattice and a renormalization of the electronic structure \cite{luo2018efficient}. This process can lower the energy of the subsequent radiative recombination, leading to the observed redshift in PL. Note that the presence of defects such as dopants can further enhance such mechanisms by introducing lattice distortions or localized defect states that act as nucleation centers for self-trapping and facilitate the stability of localized excitations. In this context, the relaxation process can be interpreted in terms of STEs \cite{luo2018efficient, xu2021toward}. 

To investigate the possibility of this self-trapping mechanism we employed constrained spin-polarized DFT (cDFT) calculations. This approach does not quantitatively capture electron-phonon coupling, but rather, is used here to qualitatively assess whether a neutral excitation can drive symmetry-breaking lattice distortions that localize in the crystal \cite{jin2024self, yu2026perspective}. In practice, local distortions ($\sim 0.1$ to $0.3$~\AA) were introduced in selected octahedra (SbCl$_6$ or InCl$_6$) and the system was structurally relaxed under different electronic constraints such as forced excited electronic occupation or finite starting magnetization. The emergence of a stable localized electronic state accompanied by a local structural distortion may be taken as a signature of the formation of STEs \cite{chen2023strong, yu2026perspective}. Finally, by comparing constrained and unconstrained relaxations, we assess whether such localized solutions are intrinsically stable or depend on the imposed initial conditions. All additional details of the cDFT approach can be found in the Materials and Methods Section and SI.

\begin{figure*}
    \centering
    \includegraphics[scale=0.49]{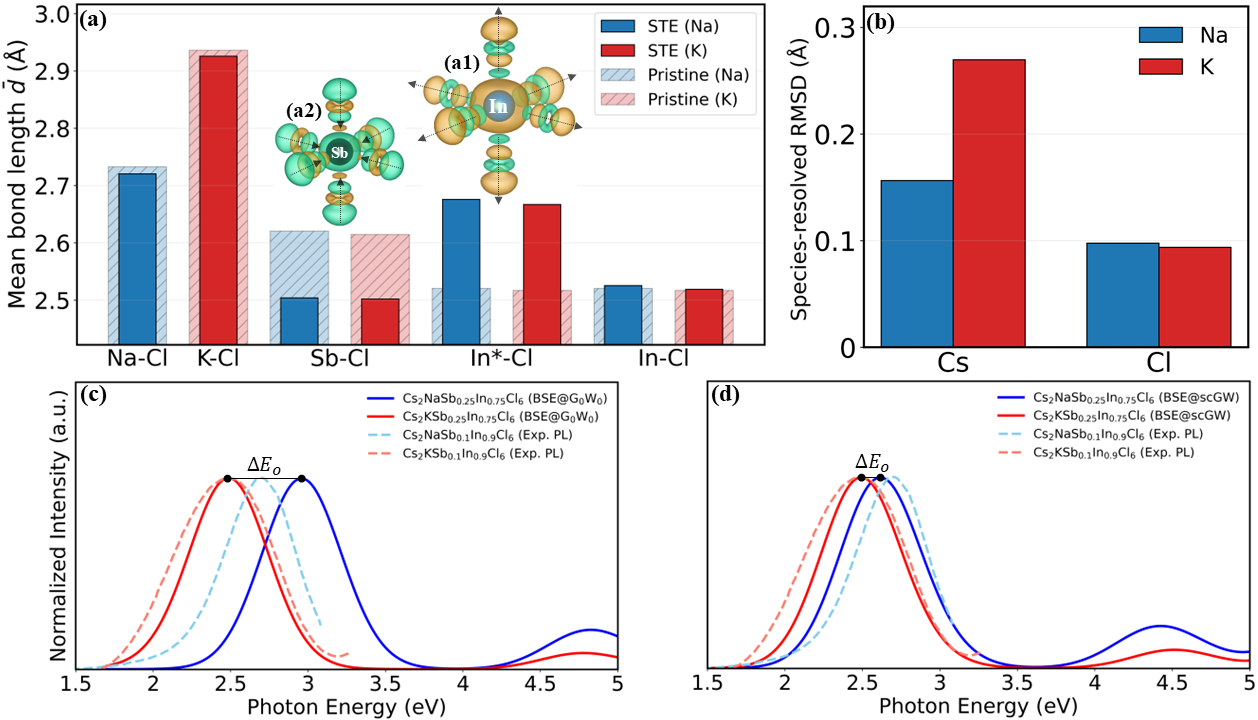}
    \caption{In the upper panels, [\textbf{a}] mean bond lengths in the pristine and constrained (STE) structure of Cs$_2$NaSb$_{0.25}$In$_{0.75}$Cl${_6}$ (blue) and Cs$_2$KSb$_{0.25}$In$_{0.75}$Cl${_6}$ (red). Graphical representation of the expansion [\textbf{a1}] and compression [\textbf{a2}] breathing modes of, respectively, In$^*$ and Sb sites. Electron accumulation (depletion) regions, obtained from the charge density difference between the pristine and STE case, are highlighted in orange (green). [\textbf{b}] Species-resolved mass-weighted RMSD $D^{(s)}_{\mathrm{RMSD}}$ of Cs$_2$NaSb$_{0.25}$In$_{0.75}$Cl${_6}$ (blue) and Cs$_2$KSb$_{0.25}$In$_{0.75}$Cl${_6}$ (red) for $s=$ Cs (0.16 and \SI{0.27}{\angstrom}) and $s=$ Cl ($\sim$ \SI{0.1}{\angstrom}) atoms (RMSD for the other atomic species is negligible). In the lower panels, linear-response optical absorption spectra, calculated at the BSE@$G_0W_0$ [\textbf{c}] and BSE@sc$GW$ [\textbf{d}], of constrained (STE) Cs$_2$NaSb$_{0.25}$In$_{0.75}$Cl${_6}$ and Cs$_2$KSb$_{0.25}$In$_{0.75}$Cl${_6}$ (solid lines) compared with the experimental photoluminescence (PL) spectra of Cs$_2$NaSb$_{0.1}$In$_{0.9}$Cl${_6}$ and Cs$_2$KSb$_{0.1}$In$_{0.9}$Cl${_6}$ (dashed lines). The difference between the optical onsets is defined as $\Delta E_\mathrm{o}$. Theoretical spectra were rigidly blueshifted to match the lowest experimental onset and a Gaussian broadening of 0.25 eV was applied. 
    \label{fig_4}}
\end{figure*}

The cDFT analysis shows a clear contrast between pristine and Sb-doped materials. In the pristine case, no robust localized state could be stabilized within the present approach. In contrast, upon Sb doping spin-polarized relaxation consistently converged to a unique triplet solution, independently of the constraints used, accompanied by a characteristic pattern of local lattice relaxation (see Fig.~\ref{fig_4}). This pattern follows a contraction of the SbCl$_6$ octahedron (Fig.~\ref{fig_4}\textit{a}2), coupled to an expansion of a single neighboring In$^*$Cl$_6$ octahedron (Fig.~\ref{fig_4}\textit{a}1), and compensating distortions of the surrounding lattice. We quantify these induced structural rearrangements by the mean bond length $\bar d = \frac{1}{6}\sum_{i=1}^{6} d_i$ (Table~S2 in the SI) of the surrounding Cl ligands (Fig.~\ref{fig_4}\textit{a}), as well as by the species-resolved, mass-weighted root-mean-square displacement (RMSD), $D_{\mathrm{RMSD}}^{(s)} = \sqrt{\sum_s m_s ||\Delta r_s||^2 / \sum_s m_s}$, where the sum runs over the atomic species $s$, $m_s$ is the atomic mass and $\Delta r_s$ is the relative atomic displacement after the relaxation. We note that the overall $D_{\mathrm{RMSD}}$ is primarily driven by Cl and Cs motions (Fig.~\ref{fig_4}\textit{b}), highlighting the soft nature of the halide lattice. Analyzing the charge density reveals that the excitation localizes over a Sb-Cl-In$^*$ cluster, with clear electron accumulation (Fig.~\ref{fig_4}\textit{a}1) and depletion (Fig.~\ref{fig_4}\textit{a}2) on the two coupled sites (In$^*$Cl$_6$ and SbCl$_6$, respectively). This pattern is consistent across both Na- and K-based systems, without clarifying the selectivity of PL peaks observed experimentally. However, a main structural difference emerges from the $D_{\mathrm{RMSD}}$ of Cs, which is substantially larger in the K-based DP ($\sim70$\%), indicative of a stronger electron–phonon coupling \cite{chen2023strong}. This behavior can be rationalized by considering the different bonding environments in the systems. In the Na-based DP, the smaller and more electronegative Na ions form stronger and more compact Na-Cl bonds, leading to a tighter lattice that constrains Cs motion. In contrast, in the K-based system, K ions exhibit lower electronegativity and larger ionic radii, resulting in weaker, more diffuse bonding and a larger lattice parameter. This provides increased free volume for Cs atoms, consistent with the larger $D_{\mathrm{RMSD}}$ observed.

We performed BSE@$G_0W_0$ calculations on the relaxed geometries obtained via the cDFT approach. The breathing modes of the SbCl$_6$ and In$^*$Cl$_6$ octahedra lead to a substantial reduction of the band gap, driven by the upward (downward) shift of the Sb-Cl hybridized valence band (In$^*$-Cl hybridized conduction band), as shown in Fig.~S5\textit{a}-\textit{b} in the SI. Consequently, optical gaps are significantly reduced (Table~\ref{tab_2}), while the difference between the optical onsets of the Na- and K-based DPs, $\Delta E_\mathrm{o}$, is enhanced due to the larger structural rearrangement in the K compound, such that $E_{o}^{\mathrm{BSE@}G_0W_0}$[K] $< E_{o}^{\mathrm{BSE@}G_0W_0}$[Na]. Absorption spectra in Fig.~\ref{fig_4}\textit{c} are in good qualitative agreement with experimental PL data up to a rigid shift of $\sim$1.7 eV and an overestimation of $\Delta E_\mathrm{o}$ (0.47 eV compared to $\sim0.2$ eV in the experiments). Motivated by this qualitative agreement, we further performed BSE calculations on top of eigenvalue-self-consistent $GW$ (BSE@sc$GW$). As shown in Fig.~\ref{fig_4}\textit{d} (see also Table~\ref{tab_2}), this approach yields a markedly improved description of the optical response: the rigid shift is reduced to $\sim0.8$ eV, and $\Delta E_\mathrm{o}$ decreases to 0.12 eV, in much closer agreement with the experiment. 

\begin{figure*}
    \centering
    \includegraphics[scale=0.49]{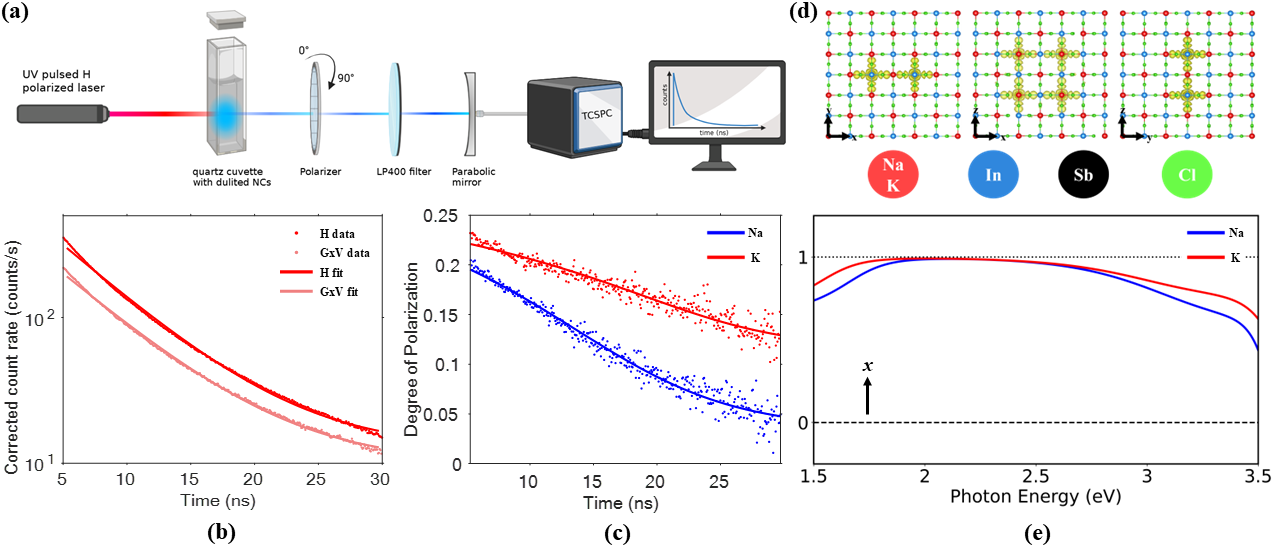}
    \caption{[\textbf{a}] Schematic illustration of the polarization-resolved measurement setup. From left to right the setup comprises a pulsed UV laser, quartz cuvette containing NCs dispersed in decane, a polarizer, a 400 nm long-pass filter, a parabolic mirror, a TCSPC detector, and an analyzer (created with BioRender.com). [\textbf{b}]  Polarization-resolved TRPL decays of Cs$_2$KSb$_{0.1}$In$_{0.9}$Cl$_6$ NCs dispersed in decane, measured under 353 nm excitation at a repetition rate of 2.069 MHz and an optical density of 0.13. The data are shown on a semi-logarithmic scale. The horizontal polarization component, $I_H$, is shown in red, and the $G$-factor-corrected vertical component, $G \times I_V$, is shown in light red. The corresponding raw data for the Na-based NCs are shown in Fig.~S12 in the SI. [\textbf{c}] Time-dependent fluorescence degree of polarization (DoP) for Cs$_2$KSb$_{0.1}$In$_{0.9}$Cl$_6$ ($H$ in red and $V \times G$ in light red) and Cs$_2$NaSb$_{0.1}$In$_{0.9}$Cl$_6$ ($H$ in blue and $V \times G$ in light blue) NCs. The black dots represent the experimental data, and the solid lines represent the corresponding fits. [\textbf{d}] Real-space representations of the first bright-exciton wavefunctions in the Na- and K-based systems, shown in the $xy$, $xz$, and $yz$ planes. Cs atoms were omitted for clarity. [\textbf{e}] DoP of optical absorption spectra, calculated at the BSE@sc$GW$, of the constrained (STE) Cs$_2$NaSb$_{0.25}$In$_{0.75}$Cl${_6}$ (blue) and Cs$_2$KSb$_{0.25}$In$_{0.75}$Cl${_6}$ (red), showing the anisotropy of the optical response along the $x$($y$)-direction.
    \label{fig_5}}
\end{figure*}

The corresponding exciton wavefunction for the lowest bright exciton (Fig.~\ref{fig_5}\textit{d}) remains strongly localized on the In$^*$ site and exhibits a pronounced anisotropy in the $xz$ plane. The dominant contributions for this excitation originate from transitions between the uppermost valence band and the lowest conduction band, primarily along the X-M direction (Fig.~S6\textit{a}–\textit{b} in the SI) in both DPs. Notably, the exciton displays an $s$-like character (first transition of the series, doubly-degenerate), in contrast to the clear $p$-like character observed in the undistorted structures. 

The anisotropic spatial feature of the excitonic wavefunction strongly suggests a polarization-dependent optical response. To verify this, we define the degree of polarization (DoP) as $[A_x-A_y]/[A_x+A_y]$, where $A_x$ ($A_y$) is the absorption for light polarized along the $x$($y$)-direction, giving the polarization direction at each photon energy. Its interpretation is straightforward: a DoP of $\pm1$ means only $x/y$-polarized absorption; a DoP of 0 means isotropic in-plane absorption. As shown in Fig.~\ref{fig_5}\textit{e} (and Fig.~S7 in the SI), both DPs are sensitive to polarization along the $x$-direction, up to $\sim$ 3.5 eV, meaning that the excitation/emission in the STE has a preferred light orientation with anisotropies predicted as high as $\sim$90\%. Moreover, the K-based DP shows a more pronounced DoP (by $\sim$13\%), consistent with the underlying structural distortions happening in the DPs, as highlighted by the RMSD. Note that emission is not guaranteed to mirror absorption anisotropy if the excited state relaxes or reorients before emission, requiring temporally resolved measurements. The STE must remain strongly symmetry-broken and oriented for the effect to be discernible, motivating single-crystal measurements excited by a linearly polarized and coherent laser source. 

We test this prediction directly with time-resolved emission decay (Fig.~\ref{fig_5}\textit{a-c}). We use an adapted version of past experiments on rotational diffusion of molecules \cite{ameloot2013fluorescence, weber1953rotational} to measure emission anisotropy decay for nanocrystals (NCs) dispersed in decane in different polarizations \cite{petrov2005timeresolved, lakowicz1999luminescence}. Decane was used to maximize the rotational diffusion time ($\sim$ 1 $\mu$s, see SI for details). A horizontally polarized pulsed laser excites the NCs with fixed polarization, preferentially exciting NCs with absorption dipoles aligned with the laser polarization.
PL emission is collected using single-photon avalanche diodes (SPADs) in two separate configurations, selected by a polarizer either parallel to the excitation polarization (horizontal, $H$), or perpendicular to it (vertical, $V$). Both polarization components were recorded as time-resolved decay traces using Time-Correlated Single Photon Counting (TCSPC) as schematically illustrated in Fig.~\ref{fig_5}a. All measurements were performed under identical laser conditions, with only the polarizer angle being modified. This measurement process allows probing whether the exciton is in a symmetry-broken self-trapped state (which should lead to strong anisotropic emission), or in a symmetric state (which should lead to unpolarized emission). Note that the surface ligands, oleic acid and oleylamine, are both achiral molecules and thus are not expected to contribute to anisotropic behavior. Representative raw data measured for each polarization are presented in Fig.~\ref{fig_5}b (additional data in Fig.~S12 in the SI).

%\begin{figure}
%    \centering
%    \includegraphics[width=0.5\linewidth]{anisotrapy figure.png}
%    \caption{[\textbf{a}] Polarization-resolved TRPL decays of Cs$_2$NaSb$_{0.1}$In$_{0.9}$Cl$_6$ nanocrystals dispersed in decane, measured under 353~nm excitation at a repetition rate of 2.069~MHz and an optical density of 0.13. The data are shown on a semilogarithmic scale, with the horizontal polarization component, $I_H$, in blue and the G-factor-corrected vertical component, $G I_V$, in red. [\textbf{b}] Time-dependent fluorescence anisotropy, $r(t)$, for Cs$_2$NaSb$_{0.1}$In$_{0.9}$Cl$_6$ nanocrystals. The experimental data are shown as dotted curves, and the fit is shown as a solid purple curve. [\textbf{c}] Polarization-resolved TRPL decays of Cs$_2$KSb$_{0.1}$In$_{0.9}$Cl$_6$ nanocrystals measured under the same conditions. The horizontal polarization component, $I_H$, is shown in blue, and the G-factor-corrected vertical component, $G I_V$, is shown in red. [\textbf{d}] Time-dependent fluorescence anisotropy, $r(t)$, for Cs$_2$KSb$_{0.1}$In$_{0.9}$Cl$_6$ nanocrystals. The experimental data are shown as dotted curves, and the fit is shown as a solid purple curve.}
%    \label{fig:placeholder}
%\end{figure}

The figure presents time-resolved PL decays of both samples after background correction. The PL profile is characteristic of DP, showing bi-exponential behavior with a short nanosecond-scale component and a longer microsecond-scale component \cite{zeng2020highly, Noculak2020}. The instrument response function (IRF) of the full measurement setup was measured, and only data sufficiently separated from the IRF-dominated region were analyzed (see Fig.~S10 in the SI). On longer timescales, in both samples, the anisotropy vanishes and PL becomes isotropic, which is expected behavior as the self-trapped excitonic system relaxes. This occurs regardless of the NCs medium (see Fig.~S13 in the SI), validating the connection to the DP nanoparticles themselves. In contrast, on shorter timescales, a strong polarization anisotropy is observed in both Cs$_2$KSb$_{0.1}$In$_{0.9}$Cl$_6$ and Cs$_2$NaSb$_{0.1}$In$_{0.9}$Cl$_6$ NCs. 
%Due to system limitations the data was measured up to 500 ns, to which we fitted the long timescale component. 
Fig.~\ref{fig_5}\textit{c} shows the time-dependent degree of polarization (DoP), calculated from the raw data and the fitted curves using the experimental notation by:

\begin{equation}
    \mathrm{DoP}(t) = \frac{I_H(t) - G I_V(t)}{I_H(t) + G I_V(t)}
    \label{eq:placeholder_label}
\end{equation}

where $I$ is the intensity of the time-resolved decay in $H$ or $V$ polarization, and $G$ is the correction factor for the polarizer-biased sensitivity (see $G$ factor calculation in Fig.~S11 in the SI).
%The absolute difference in maximum anisotropy between the K- and Na-based NCs is 2\%, similar to the theoretical prediction we introduced. The K-based NCs exhibit higher anisotropy than the Na- based NCs- $r_{\max}=0.1677 \pm 0.003$ and $r_{\max}=0.146 \pm 0.002$, respectively.
The K-based NCs exhibit a higher maximum anisotropy than the Na-based NCs, with $\mathrm{DoP}_{\max}=0.23 \pm 0.004$ and $\mathrm{DoP}_{\max}=0.2 \pm 0.003$, respectively.
The absolute difference, $\Delta \mathrm{DoP}_{\max}\sim0.03$, corresponds to a 15\% relative increase, which is remarkably comparable to the theoretical prediction from MBPT in the STE states. 

Note that the absolute values of anisotropy measured experimentally are much lower compared to the theoretical prediction, which we expect due to the mesoscopic nature of the samples that are not uniformly aligned (the predicted value is for a single crystal sample). Moreover, the experimental anisotropy is extracted from time-resolved measurements rather than steady-state emission as in the calculation, which also leads to discrepancies. Lastly, the measured anisotropy may be influenced by several additional factors, ranging from surface-related effects to rotational dynamics. Despite these points, the observation of such strong anisotropic PL from mesoscopic NCs in solution implies that the anisotropy from a single crystal should be extremely high (in accordance with our theory), which constitutes a strong spectral signature of STE physics.
%Although the PL anisotropy does not directly measure local distortion of the cubic symmetry as the STE energy forms, we believe it is a clear indication of the process characteristics. It is clear that, both theoretically and experimentally, there is an anisotropic behavior which is more pronounced for K-based DP.

%To compare the values obtained with the theoretical results, we use time-resolved fluorescence anisotropy and the degree of linear polarization (DOLP) relationship:
%\begin{equation}
 %   \mathrm{DOLP}(t) = \frac{3r(t)}{2 + r(t)}
  %  \label{eq:placeholder_label}
%\end{equation}
%The maximum DOLP is $0.23 \pm 0.004$ and $0.20 \pm 0.003$ for K-based NCs and Na based, respectively. 

Further analyzing the experiment, we find that the anisotropy persists for a longer duration in K-based NCs (65 ns) than in Na-based NCs (45 ns) (see Fig.~S14 in the SI), which can be used as a separate signature of STE composition and lattice structure. This is consistent with our calculations that show the K-based compound exhibits a softer lattice and undergoes greater structural distortion upon excitation (see Fig.~\ref{fig_4}). 
%This anisotropy decay is unlikely to originate from rotational diffusion, as the estimated rotational correlation time remains on the microsecond timescale even under the most conservative assumptions (see Fig. SIx).

%To validate the anisotropy measurements, additional samples were measured at different NC optical densities and under different laser conditions. A dependence on the optical density was observed, with lower optical density resulting in a higher maximum anisotropy. This trend may be attributed to concentration-dependent effects such as light scattering, reabsorption, energy transfer, or aggregation. Moreover, the depolarization time point (H and V polarization decays present isotropic behavior) exceeds for the low concentration sample, which can also originate from the mentioned effect (see Fig. SIX). Furthermore, repeated H- and V-polarized measurements were performed on K-based NCs to confirm the reproducibility of the observed anisotropic response and to rule out measurement-induced particle degradation (see SI-X).

Altogether, our results suggest a full physical picture of STEs in DPs. Doping introduces strongly hybridized in-gap states, delocalizing the excitonic wavefunction and promoting lattice-assisted localization of excited states. In turn, this leads to a red-shifted PL. The choice of B cation tunes exciton-phonon couplings, leading to chemically tunable red-shift in PL. Finally, PL anisotropy arises as a signature of STEs and is sensitive to the composition and STE-induced lattice distortion. It provides indirect access to the excitonic wavefunction itself and its spatial distribution.

\begin{table}[h!]
    \centering
    \caption{Estimate of the lowest electronic gap at the G$_0$W$_0$ ($E^{G_0W_0}$) and self-consistent $GW$ level ($E^{scGW}$), of the Sb-doped constrained (STE) structures. Also, the optical gap ($E_{o}^{\mathrm{BSE@}G_0W_0}$, $E_{o}^{\mathrm{BSE@sc}GW}$) and the binding energy ($E_{B}^{\mathrm{BSE@}G_0W_0}$, $E_{B}^{\mathrm{BSE@sc}GW}$) of the first bright exciton at the BSE level are displayed. For indirect electronic gaps, the lowest direct electronic gaps are shown in round brackets. $\Delta E_\mathrm{o}$ represents the difference between the optical threshold of the two systems (reference experimental value: $\approx 0.2$). All values are expressed in eV.}
    \label{tab_2}
    \small
    \renewcommand{\arraystretch}{1.3}
    \begin{tabular}{l|c|c|c}
        \toprule
        & \textbf{Cs$_{\mathbf{2}}$NaSb$_{\mathbf{0.25}}$In$_{\mathbf{0.75}}$Cl$_{\mathbf{6}}$}
        & \textbf{Cs$_{\mathbf{2}}$KSb$_{\mathbf{0.25}}$In$_{\mathbf{0.75}}$Cl$_{\mathbf{6}}$} & $\bm{\Delta E_\mathrm{o}}$ \\
        \midrule
        \textbf{E$^{\mathbf{G_0W_0}}$} 
            & 1.61 (1.70) & 0.98 (1.19) & \\
        \textbf{E$_\mathbf{o}^{\mathbf{BSE@G_0W_0}}$} 
            & 1.25 & 0.78 & 0.47 \\
        \textbf{E$_\mathbf{B}^{\mathbf{BSE@G_0W_0}}$} 
            & 0.45 & 0.41 & \\
        \midrule
        \textbf{E$^{\mathbf{scGW}}$} 
            & 2.29 (2.34) & 2.00 (2.14) & \\
        \textbf{E$_\mathbf{o}^{\mathbf{BSE@scGW}}$} 
            & 1.80 & 1.68 & 0.12 \\
        \textbf{E$_\mathbf{B}^{\mathbf{BSE@scGW}}$} 
            & 0.54 & 0.46 & \\
        \bottomrule
    \end{tabular}
\end{table}

\section{Conclusions}

In summary, we carried out a combined theoretical and experimental investigation of PL and absorption in lead-free (Sb-doped) Cs$_2$K$_x$Na$_{1-x}$InCl$_6$ DPs.
Our measurements reveal clear signatures of self-trapped excitonic states in red-shifted PL relative to optical absorption. The PL is strongly sensitive to crystal composition, shifting by as much as $\sim$0.2 eV across alloys, whereas the optical absorption remains comparatively composition independent.
Through a detailed \textit{ab initio} analysis, we confirm the self-trapped nature of the excitons, which arises from strong electron-phonon coupling and reduces the optical gap of the excited system. Our simulations uncover the microscopic mechanisms governing exciton formation and the chemical tunability of the PL peaks. Specifically, we show that Na/K substitution on the B site modifies the self-trapped lattice distortion around Cs ions, inducing stronger energy shifts of the dopant-derived orbitals that form the exciton, thereby red-shifting PL. We further revealed that Sb doping delocalizes the exciton wavefunction through hybridization with in-gap flat states, likely enhancing the oscillator strength of the transition. Interestingly, the STE state is stable only in the doped system, suggesting that exciton-lattice coupling depends sensitively on the delocalized character of the excitation. Lastly, we predicted a pronounced optical anisotropy in the PL, which was validated in time-resolved measurements. The anisotropy in PL directly probes the spatial structure of the STEs, which is similarly tunable with alloying, providing direct optical probing of excitonic wavefunctions.

Overall, our results shed light on the microscopic mechanisms underlying STEs in lead-free DPs, with particular emphasis on the distinct roles of dopants and alloying in tuning the excitonic state and exciton–lattice coupling. This understanding could enable the selective engineering of emission and absorption properties in emerging crystal systems and support the development of lead-free DPs for optoelectronic applications. Moreover, we expect that optical anisotropy measurements could form a novel approach for probing excitonic wavefunctions, especially in self-trapped states where pump-probe schemes can directly track the coupled electron-hole-lattice dynamics in time.  
Looking ahead, similar analyses may reveal even more intricate behavior in nanoscale systems and in materials with surface effects. Future work should focus on experimentally probing the predicted variation in STE bond lengths and directly validate the dynamic STE formation process. 
%This could be achieved using advanced time-resolved techniques such as ultrafast electron microscopy, electron diffraction, or electron energy-loss spectroscopy, which can resolve structural and electronic dynamics on the relevant timescales.

\section{Acknowledgments}
ON gratefully acknowledges the support of the Young Faculty Award from the National Quantum Science and Technology program of Israel’s Council of Higher Education Planning and Budgeting Committee, as well as Nevet grants from the Technion RBNI and Helen-Diller Quantum centers. SG gratefully acknowledges Friedhelm Bechstedt, Olivia Pulci, Maurizia Palummo and Giacomo Giorgi for the valuable comments and suggestions. SG also acknowledges the CINECA award under the ISCRA initiative, for the availability of high performance computing resources and support. UDG acknowledges support from the Marie Skłodowska-Curie Doctoral Networks TIMES (grant No. 101118915) and SPARKLE (grant No. 101169225), and from the Italian Ministry of University and Research (MUR) under PRIN 2022 (grant No. 2022PX279E003). YB acknowledges the support for this project by the European Union’s Horizon 2020 research and innovation program under grant agreement No. 949682 (ERC). This work was supported by the European Research Council (ERC-2024-SyG-101167294; UnMySt), the Cluster of Excellence Advanced Imaging of Matter (AIM), Grupos Consolidados y Alto Rendimiento UPV/EHU, Gobierno Vasco (IT1453-22). 
We acknowledge support from the Max Planck - New York City Center for Non-Equilibrium Quantum Phenomena.
OZ and YB thank Kobi Cohen for the assistance with the anisotropic measurement setup.

\section{Materials and methods}

\subsection{Experimental details}

\noindent \textbf{Materials.} Antimony (III) acetate (99.9\%, Aldrich), benzoyl chloride (99.9\%, Alfa Aesar), cesium carbonate (99.9\%, Aldrich), hexane (A.R. Aldrich or 99.9\%, Fischer Scientific), indium (III) acetate (99.9\%, Aldrich), octadecene (90\%, Aldrich), oleic acid (90\%, Aldrich), oleylamine (70\%, Aldrich), silver acetate (99.9\%, Aldrich), sodium acetate (99.9\%, Aldrich). All chemicals were used as purchased without additional purification.

\noindent \textbf{Synthesis of 0.5M Cs-oleate precursor.} The Cs-oleate solution was prepared following the approach of Locardi et al. \cite{locardi2019emissive} 1.63 g (5 mmol) of Cs$_2$CO$_3$, 20 mL (63.37 mmol) of oleic acid, and a stirring bar were inserted into a 50 mL three-necked round bottom flask. The flask was plugged into a Schlenk line and degassed under vacuum and 100\textsuperscript{o}C for 30 minutes or until there were no bubbles. Then, the temperature was raised to 150\textsuperscript{o}C under nitrogen and stirred for 3 hours. The product was a clear yellow solution.

\noindent \textbf{Synthesis of 0.25M Sb-oleate, 0.25M In-oleate , 0.5M Na-oleate, 0.5M K-oleate precursors.} 0.7472 g (2.5 mmol) of Sb(ac)$_3$ or 0.7299 g (2.5 mmol) of In(ac)$_3$ or 0.4102 g (5 mmol) of Na(ac) or 0.4907 g (5 mmol) with 10 mL (31.7 mmol) of oleic acid and a stirring bar were inserted into a 25 mL three-neck flask. The reaction was stirred for 45 minutes at 90°C under vacuum and then filled with N$_2$ gas under heat for 10 more minutes. The Sb-oleate and Na-oleate products were slightly yellow, clear solutions. 

\noindent \textbf{Synthesis of Cs$_2$Na$_{1-x}$K$_x$In$_{0.9}$Sb$_{0.1}$Cl$6$ NCs.} Cs$_2$Na$_{1-x}$K$_x$In$_{0.9}$Sb$_{0.1}$Cl$_6$ NCs were synthesized using the procedures developed by Locardi et al. \cite{locardi2019emissive} with several modifications. 0.1 mL of 0.25M Sb-Oleate solution (0.025 mmol), 1 mL of 0.5M Cs-oleate solution (0.5 mmol), 0.9 mL of 0.25M In-oleate solution (0.225 mmol), 0.5 mL oleylamine (1.52 mmol), 4.5 mL of ODE, different ratios of 0.5M Na-oleate and 0.5M K-oleate solution between 0-0.48 mL (0-0.24 mmol) and a stirring bar were inserted into a 25 mL three-neck flask. The reaction was stirred for 45 minutes at 110°C under vacuum and then filled with N$_2$ gas under heat for 10 more minutes. Then, 200 µL of Bz-Cl (1.72 mmol) was swiftly injected. The reaction was stirred for another 5 seconds and then cooled in a cold water bath. The solution was then centrifuged, first at 7000 rpm for 10 minutes, and the precipitate was redispersed in 5 mL hexane and then centrifuged at 3500 rpm for 5 minutes. The solution was transferred to a new tube and centrifuged again at 7000 rpm for 10 minutes and separated from the residue.

\noindent \textbf{Optical characterizations.} Absorption measurements were performed using a Gen5 Synergy H1 hybrid multi-mode reader spectrophotometer with 96-well plates. Emission and photoluminescence excitation (PLE) measurements were carried out using an Edinburgh Instruments FLS1000 photoluminescence spectrometer with excitation provided by a xenon lamp. All measurements were performed on the products dispersed in a hexane emulsion, with clean hexane used as a blank reference. Spectra were recorded with a step size of 1 nm. For the Synergy measurements, the detector gain was set to either 50 or 100, depending on the specific product and sample concentration. 

\noindent \textbf{Time-Resolved PL Decay Anisotropy Measurements}. Time-resolved photoluminescence measurements were performed on NC dispersions in decane using quartz cuvettes. The samples were diluted to an optical density of 0.09-0.13. Excitation was performed using a pulsed 353 nm laser with a pulse width of 3 ps and a repetition rate of 1-2 MHz and average power of 2.875-5.75 mW, respectively.  For polarization-resolved measurements, the excitation polarization was fixed, while the emission was analyzed using a linear polarizer oriented parallel and perpendicular to the excitation polarization. The detected emission was filtered using a 400 nm long-pass filter before reaching an HPM-100-07 hybrid photodetector module. Photon arrival times were recorded using PicoHarp 330 TCSPC software, with an overall system response time of 20 ps. The system operates in a start–stop configuration, where the laser pulse serves as the START signal, and the detection of the corresponding photon at the detector provides the STOP signal. 
The polarization-dependent detection efficiency was corrected using a G-factor determined from an isotropic reference sample. Each sample was measured in horizontal (H) and vertical (V) polarization. More details on the measurements and reference sample are available in the SI. 

\noindent \textbf{X-ray diffraction (XRD) characterization.} The NC solution in hexane was centrifuged at 12000 rpm for 10 minutes, and the residue was drop-cast onto a rectangular micro slide glass substrate (76 mm $\times$ 26 mm). Measurements were taken using a Rigaku Smart-Lab 9 kW high-resolution X-ray diffractometer, equipped with a rotating anode X-ray source. We used a \SI{1.54}{\angstrom} (Cu K$\alpha$) wavelength and a 2$\theta$ range of 10\textsuperscript{o}-90\textsuperscript{o}.

\noindent \textbf{TEM characterizations.} A drop of dilute NC solution in hexane was cast onto a TEM grid - carbon film only on 300 mesh copper, and observed in TEM mode using a FEI/Thermo-Fisher Tecnai G$^2$ T20 with LaB$_6$ electron source operated at an accelerating voltage of 200 keV.

\noindent \textbf{High-angle annular dark-field scanning transmission electron microscopy (HAADF-STEM).} A TEM grid was prepared the same as for the TEM characterization and observed in HAADF-STEM mode using a FEI/Thermo-Fisher double corrected 60-300 Titan Themis FEG-S/TEM operated at an accelerating voltage of 200 keV.

\noindent \textbf{Scanning electron microscopy (SEM) and energy-dispersive X-ray spectroscopy (EDS).} The NC solution in hexane was centrifuged at 12000 rpm for 10 minutes, and the residue was drop-cast onto a p-type silicon wafer slice (10$\times$10 mm$^2$). SEM micrographs were taken using HR-SEM Zeiss Ultra-Plus microscope at an acceleration voltage of 4 kV. For EDS measurements, the samples were positioned at a working distance of 7 mm and analyzed using an acceleration voltage of 7 kV. 

\subsection{Theoretical details}

\noindent Ground-state properties were calculated within the framework of density functional theory (DFT) using the \textsc{Quantum ESPRESSO} (QE) code \cite{giannozzi2009quantum}. Plane-wave norm-conserving pseudopotentials, including non-linear core corrections, were taken from the PseudoDojo repository \cite{PseudoDojo}, both in their standard form (without semi-core states) and in the stringent version (including semi-core states). For the exchange–correlation (XC) functional, we employed the generalized gradient approximation (GGA) in the parametrization of Perdew–Burke–Ernzerhof (PBE) \cite{PBE} and its solid-state revision PBEsol \cite{PBEsol}. Fully relativistic pseudopotentials were also used to incorporate spin–orbit coupling (SOC). \\
\noindent Full structural relaxation of the cubic 40-atom unit cell was performed while enforcing symmetry, and considered converged once the maximum residual ionic force dropped below $10^{-8}$ Ry/Bohr. The converged ground-state calculations employed a kinetic energy cutoff of 80 Ry and a uniform Monkhorst-Pack \cite{monkhorst1976} $\mathbf{k}$-point grid of $4\times4\times4$. \\
\noindent Many-body perturbation theory (MBPT) calculations were subsequently performed with the \textsc{Yambo} code \cite{marini2009yambo}. Quasiparticle (QP) corrections to the DFT Kohn-Sham (KS) electronic states were obtained within the $G_0W_0$ and eigenvalue self-consistent $GW$ (sc$GW$) schemes. Excitonic effects in the optical spectra were included by solving the Bethe–Salpeter equation (BSE) \cite{BSE1, BSE2, BSE3, BSE4}, considering only direct transitions at $q=0$. \\
\noindent The $GW$ calculations employed the plasmon-pole approximation (PPA) for the frequency dependence of the dielectric function, as implemented in \textsc{Yambo}. For the BSE Hamiltonian, a total of 68 (70) occupied and 20 (18) unoccupied states, together with a $\mathbf{k}$-point grid of $4\times4\times4$, were considered upon convergence for the pristine (Sb-doped) systems.

\bibliography{bibliography}

@article{Noculak2020,
author = {Noculak, Agnieszka and Morad, Viktoriia and McCall, Kyle M. and Yakunin, Sergii and Shynkarenko, Yevhen and W{\"o}rle, Michael and Kovalenko, Maksym V.},
title = {Bright Blue and Green Luminescence of Sb(III) in Double Perovskite Cs2MInCl6 (M = Na, K) Matrices},
journal = {Chemistry of Materials},
volume = {32},
number = {12},
pages = {5118-5124},
year = {2020},
doi = {10.1021/acs.chemmater.0c01004},
    note ={PMID: 32595266},
}

@article{Gill2025,
author = {Gill, D and Shallcross, S and Chen, W and Dewhurst, J K and Sharma, S},
doi = {10.1021/acs.nanolett.5c01871},
issn = {1530-6984},
journal = {Nano Letters},
month = {aug},
number = {33},
pages = {12439--12445},
publisher = {American Chemical Society},
title = {{Coupled Femtoexcitons, Free Carriers, and Light}},
url = {https://doi.org/10.1021/acs.nanolett.5c01871},
volume = {25},
year = {2025}
}

@article{Chen2026,
  title={Attosecond three-stage formation and coherent exciton dynamics in a two-dimensional material under strong field},
  author={Chen, Qing and Chen, Daqiang and Wang, Chenyu and Bai, Yunfei and Lian, Chao and Xu, Zhe and Guo, Haizhong and Wang, Enge and Meng, Sheng},
  journal={Light: Science \& Applications},
  volume={15},
  number={1},
  pages={217},
  year={2026},
  publisher={Nature Publishing Group UK London}
}

@article{Anderson1993,
author = {Anderson, Mark T and Greenwood, Kevin B and Taylor, Gregg A and Poeppelmeier, Kenneth R},
doi = {https://doi.org/10.1016/0079-6786(93)90004-B},
issn = {0079-6786},
journal = {Progress in Solid State Chemistry},
number = {3},
pages = {197--233},
title = {{B-cation arrangements in double perovskites}},
url = {https://www.sciencedirect.com/science/article/pii/007967869390004B},
volume = {22},
year = {1993}
}

@article{Muscarella2022,
author = {Muscarella, Loreta A and Hutter, Eline M},
doi = {10.1021/acsenergylett.2c00811},
journal = {ACS Energy Letters},
month = {jun},
number = {6},
pages = {2128--2135},
publisher = {American Chemical Society},
title = {{Halide Double-Perovskite Semiconductors beyond Photovoltaics}},
url = {https://doi.org/10.1021/acsenergylett.2c00811},
volume = {7},
year = {2022}
}

@article{zhou2019manipulation,
  title={Manipulation of Bi3+/In3+ transmutation and Mn2+-doping effect on the structure and optical properties of double perovskite Cs2NaBi1-xInxCl6},
  author={Zhou, Jun and Rong, Ximing and Zhang, Peng and Molokeev, Maxim S and Wei, Peijia and Liu, Quanlin and Zhang, Xiuwen and Xia, Zhiguo},
  journal={Advanced Optical Materials},
  volume={7},
  number={8},
  pages={1801435},
  year={2019},
  publisher={Wiley Online Library}
}

@article{gray2020high,
  title={High-efficiency blue photoluminescence in the Cs 2 NaInCl 6: Sb 3+ double perovskite phosphor},
  author={Gray, Matthew B and Hariyani, Shruti and Strom, T Amanda and Majher, Jackson D and Brgoch, Jakoah and Woodward, Patrick M},
  journal={Journal of Materials Chemistry C},
  volume={8},
  number={20},
  pages={6797--6803},
  year={2020},
  publisher={Royal Society of Chemistry}
}

@article{jeevaraj2024extrinsic,
  title={Extrinsic electronic states to tune the luminescence and bonding nature of Cs2NaInCl6 double perovskite},
  author={Jeevaraj, M and Sivaganesh, D and Saravanakumar, S and Bahadur, S Asath and Sudhahar, S and Kumar, M Krishna},
  journal={Materials Chemistry and Physics},
  volume={311},
  pages={128569},
  year={2024},
  publisher={Elsevier}
}

@article{rameshkumar2023exploring,
  title={Exploring the structural and optical properties of Eu3+, Sb3+ co-doping effect on Cs2NaInCl6 double perovskites for LEDs applications},
  author={Rameshkumar, S},
  journal={Solid State Communications},
  volume={376},
  pages={115375},
  year={2023},
  publisher={Elsevier}
}

@article{guedira1988phase,
  title={Phase transition in a chloro-elpasolite, Cs2KInCl6},
  author={Guedira, T and Wignacourt, JP and Drache, M and Lorriaux-Rubbens, A and Wallart, F},
  journal={Phase Transitions},
  volume={13},
  number={1-4},
  pages={81--85},
  year={1988},
  publisher={Taylor \& Francis}
}

@article{meng2017parity,
  title={Parity-forbidden transitions and their impact on the optical absorption properties of lead-free metal halide perovskites and double perovskites},
  author={Meng, Weiwei and Wang, Xiaoming and Xiao, Zewen and Wang, Jianbo and Mitzi, David B and Yan, Yanfa},
  journal={The journal of physical chemistry letters},
  volume={8},
  number={13},
  pages={2999--3007},
  year={2017},
  publisher={ACS Publications}
}

@article{khan2021elastic,
  title={Elastic and optoelectronic properties of Cs2NaMCl6 (M= in, Tl, Sb, Bi)},
  author={Khan, Imad and Haq, Izaz Ul and Akbar, Ali and Ali, Zahid and Ahmad, Iftikhar and others},
  journal={Journal of Electronic Materials},
  volume={50},
  number={2},
  pages={456--466},
  year={2021},
  publisher={Springer Nature BV}
}

@article{leppert2024excitons,
  title={Excitons in metal-halide perovskites from first-principles many-body perturbation theory},
  author={Leppert, Linn},
  journal={The Journal of chemical physics},
  volume={160},
  number={5},
  year={2024},
  publisher={AIP Publishing}
}

@article{shakeel2024dft,
  title={DFT analysis of elastic and optoelectronic properties of Cs2NaXCl6 (X= In, La, Sc, Y) double perovskite compounds},
  author={Shakeel, Shakeel and Song, Peng and Shah, Syed Hatim and Zada, Zeshan and Huang, Taihong and Laref, Amel and Hakimi, Nadimullah and Faizan, Muhammad},
  journal={Materials Chemistry and Physics},
  volume={324},
  pages={129683},
  year={2024},
  publisher={Elsevier}
}

@book{grosso2013solid,
  title={Solid state physics},
  author={Grosso, Giuseppe and Parravicini, Giuseppe Pastori},
  year={2013},
  publisher={Academic press}
}

@article{giannozzi2009quantum,
  title={QUANTUM ESPRESSO: a modular and open-source software project for quantumsimulations of materials},
  author={Giannozzi, Paolo and Baroni, Stefano and Bonini, Nicola and Calandra, Matteo and Car, Roberto and Cavazzoni, Carlo and Ceresoli, Davide and Chiarotti, Guido L and Cococcioni, Matteo and Dabo, Ismaila and others},
  journal={Journal of physics: Condensed matter},
  volume={21},
  number={39},
  pages={395502},
  year={2009},
  publisher={IOP Publishing}
}

@article{marini2009yambo,
  title={Yambo: an ab initio tool for excited state calculations},
  author={Marini, Andrea and Hogan, Conor and Gr{\"u}ning, Myrta and Varsano, Daniele},
  journal={Computer Physics Communications},
  volume={180},
  number={8},
  pages={1392--1403},
  year={2009},
  publisher={Elsevier}
}

@article{PseudoDojo,
title = "The PseudoDojo: Training and grading a 85 element optimized norm-conserving pseudopotential table",
journal = "Computer Physics Communications",
volume = "226",
pages = "39 - 54",
year = "2018",
issn = "0010-4655",
doi = "https://doi.org/10.1016/j.cpc.2018.01.012",
author = "M.J. van Setten and M. Giantomassi and E. Bousquet and M.J. Verstraete and D.R. Hamann and X. Gonze and G.-M. Rignanese",
note={\url{http://www.pseudo-dojo.org/}}
}

@article{PBEsol,
  title={Restoring the density-gradient expansion for exchange in solids and surfaces},
  author={Perdew, John P and Ruzsinszky, Adrienn and Csonka, G{\'a}bor I and Vydrov, Oleg A and Scuseria, Gustavo E and Constantin, Lucian A and Zhou, Xiaolan and Burke, Kieron},
  journal={Physical review letters},
  volume={100},
  number={13},
  pages={136406},
  year={2008},
  publisher={APS}
}

@article{PBE,
  title = {Generalized Gradient Approximation Made Simple},
  author = {Perdew, John P. and Burke, Kieron and Ernzerhof, Matthias},
  journal = {Phys. Rev. Lett.},
  volume = {77},
  issue = {18},
  pages = {3865--3868},
  numpages = {0},
  year = {1996},
  month = {Oct},
  publisher = {American Physical Society},
}

@article{monkhorst1976,
  title={Special points for Brillouin-zone integrations},
  author={Monkhorst, Hendrik J and Pack, James D},
  journal={Physical review B},
  volume={13},
  number={12},
  pages={5188},
  year={1976},
  publisher={APS}
}

@article{BSE1,
  title = {Electronic excitations: density-functional versus many-body Green's-function approaches},
  author = {Onida, Giovanni and Reining, Lucia and Rubio, Angel},
  journal = {Rev. Mod. Phys.},
  volume = {74},
  issue = {2},
  pages = {601--659},
  numpages = {0},
  year = {2002},
  month = {Jun},
  publisher = {American Physical Society},
}

@article{BSE2,
  title = {Dynamical aspects of correlation corrections in a covalent crystal},
  author = {Strinati, G. and Mattausch, H. J. and Hanke, W.},
  journal = {Phys. Rev. B},
  volume = {25},
  issue = {4},
  pages = {2867--2888},
  numpages = {0},
  year = {1982},
  month = {Feb},
  publisher = {American Physical Society},
}

@article{BSE3,
  title = {Many-particle effects in the optical spectrum of a semiconductor},
  author = {Hanke, W. and Sham, L. J.},
  journal = {Phys. Rev. B},
  volume = {21},
  issue = {10},
  pages = {4656--4673},
  numpages = {0},
  year = {1980},
  month = {May},
  publisher = {American Physical Society},
}

@article{BSE4,
  title = {Dielectric Response in the Wannier Representation: Application to the Optical Spectrum of Diamond},
  author = {Hanke, W. and Sham, L. J.},
  journal = {Phys. Rev. Lett.},
  volume = {33},
  issue = {10},
  pages = {582--585},
  numpages = {0},
  year = {1974},
  month = {Sep},
  publisher = {American Physical Society},
}

@article{shannon1976revised,
  title={Revised effective ionic radii and systematic studies of interatomic distances in halides and chalcogenides},
  author={Shannon, Robert D},
  journal={Foundations of Crystallography},
  volume={32},
  number={5},
  pages={751--767},
  year={1976},
  publisher={International Union of Crystallography}
}

@article{tran2017designing,
  title={Designing indirect--direct bandgap transitions in double perovskites},
  author={Tran, T Thao and Panella, Jessica R and Chamorro, Juan R and Morey, Jennifer R and McQueen, Tyrel M},
  journal={Materials Horizons},
  volume={4},
  number={4},
  pages={688--693},
  year={2017},
  publisher={Royal Society of Chemistry}
}

@article{biega2023chemical,
  title={Chemical mapping of excitons in halide double perovskites},
  author={Biega, Raisa-Ioana and Chen, Yinan and Filip, Marina R and Leppert, Linn},
  journal={Nano letters},
  volume={23},
  number={17},
  pages={8155--8161},
  year={2023},
  publisher={ACS Publications}
}

@article{luo2018cs2agincl6,
  title={Cs2AgInCl6 double perovskite single crystals: parity forbidden transitions and their application for sensitive and fast UV photodetectors},
  author={Luo, Jiajun and Li, Shunran and Wu, Haodi and Zhou, Ying and Li, Yang and Liu, Jing and Li, Jinghui and Li, Kanghua and Yi, Fei and Niu, Guangda and others},
  journal={Acs Photonics},
  volume={5},
  number={2},
  pages={398--405},
  year={2018},
  publisher={ACS Publications}
}

@article{zeng2020highly,
  title={Highly efficient blue emission from self-trapped excitons in stable Sb3+-doped Cs2NaInCl6 double perovskites},
  author={Zeng, Ruosheng and Zhang, Leilei and Xue, Yang and Ke, Bao and Zhao, Zhuang and Huang, Dan and Wei, Qilin and Zhou, Weichang and Zou, Bingsuo},
  journal={The Journal of Physical Chemistry Letters},
  volume={11},
  number={6},
  pages={2053--2061},
  year={2020},
  publisher={ACS Publications}
}

@article{sangalli2019many,
  title={Many-body perturbation theory calculations using the yambo code},
  author={Sangalli, Davide and Ferretti, Andrea and Miranda, Henrique and Attaccalite, Claudio and Marri, Ivan and Cannuccia, Elena and Melo, Pedro and Marsili, Margherita and Paleari, Fulvio and Marrazzo, Antimo and others},
  journal={Journal of physics: Condensed matter},
  volume={31},
  number={32},
  pages={325902},
  year={2019},
  publisher={IOP Publishing}
}

@article{hedin1965new,
  title={New method for calculating the one-particle Green's function with application to the electron-gas problem},
  author={Hedin, Lars},
  journal={Physical Review},
  volume={139},
  number={3A},
  pages={A796},
  year={1965},
  publisher={APS}
}

@article{jin2024self,
  title={Self-trapped excitons in metal-halide perovskites investigated by time-dependent density functional theory},
  author={Jin, Yu and Rusishvili, Mariami and Govoni, Marco and Galli, Giulia},
  journal={The Journal of Physical Chemistry Letters},
  volume={15},
  number={12},
  pages={3229--3237},
  year={2024},
  publisher={ACS Publications}
}

@article{chen2023strong,
  title={Strong Electron--Phonon Coupling Induced Self-Trapped Excitons in Double Halide Perovskites},
  author={Chen, Baian and Chen, Rui and Huang, Bolong},
  journal={Advanced Energy and Sustainability Research},
  volume={4},
  number={9},
  pages={2300018},
  year={2023},
  publisher={Wiley Online Library}
}

@article{yu2026perspective,
  title={Perspective on Understanding the Origin of Trapped Exciton in Lead-Halide Perovskite with Broadband Emission},
  author={Yu, Shaohua and Xu, Jin and Zeng, Weigang and Xiao, Zhiyuan and Chen, Xueyuan},
  journal={Advanced Optical Materials},
  volume={14},
  number={2},
  pages={e02683},
  year={2026},
  publisher={Wiley Online Library}
}

@article{xu2021toward,
  title={Toward a general understanding of exciton self-trapping in metal halide perovskites},
  author={Xu, Zhengwei and Jiang, Xingxing and Cai, Hua-peng and Chen, Keqiu and Yao, Xiaolong and Feng, Yexin},
  journal={The Journal of Physical Chemistry Letters},
  volume={12},
  number={43},
  pages={10472--10478},
  year={2021},
  publisher={ACS Publications}
}

@article{zhu2021sb,
  title={Sb-doped metal halide nanocrystals: a 0D versus 3D comparison},
  author={Zhu, Dongxu and Zaffalon, Matteo L and Zito, Juliette and Cova, Francesca and Meinardi, Francesco and De Trizio, Luca and Infante, Ivan and Brovelli, Sergio and Manna, Liberato},
  journal={ACS Energy Letters},
  volume={6},
  number={6},
  pages={2283--2292},
  year={2021},
  publisher={ACS Publications}
}

@article{luo2018efficient,
  title={Efficient and stable emission of warm-white light from lead-free halide double perovskites},
  author={Luo, Jiajun and Wang, Xiaoming and Li, Shunran and Liu, Jing and Guo, Yueming and Niu, Guangda and Yao, Li and Fu, Yuhao and Gao, Liang and Dong, Qingshun and others},
  journal={Nature},
  volume={563},
  number={7732},
  pages={541--545},
  year={2018},
  publisher={Nature Publishing Group UK London}
}

@article{wang2019atomistic,
  title={Atomistic mechanism of broadband emission in metal halide perovskites},
  author={Wang, Xiaoming and Meng, Weiwei and Liao, Weiqiang and Wang, Jianbo and Xiong, Ren-Gen and Yan, Yanfa},
  journal={The Journal of Physical Chemistry Letters},
  volume={10},
  number={3},
  pages={501--506},
  year={2019},
  publisher={ACS Publications}
}

@article{Slavney2016,
  author  = {Adam H. Slavney and Te Hu and Aaron M. Lindenberg and Hemamala I. Karunadasa},
  title   = {A Bismuth-Halide Double Perovskite with Long Carrier Recombination Lifetime for Photovoltaic Applications},
  journal = {Journal of the American Chemical Society},
  year    = {2016},
  volume  = {138},
  number  = {7},
  pages   = {2138--2141},
  doi     = {10.1021/jacs.5b13294}
}

@article{Nair2022,
  author  = {Sindhu S. Nair and Lucky Krishnia and Alex Trukhanov and Preeti Thakur and Atul Thakur},
  title   = {Prospect of Double Perovskite over Conventional Perovskite in Photovoltaic Applications},
  journal = {Ceramics International},
  year    = {2022},
  volume  = {48},
  number  = {23},
  pages   = {34128--34147},
  doi     = {10.1016/j.ceramint.2022.08.184}
}

@article{Chu2019,
  author  = {L. Chu and W. Ahmad and W. Liu and others},
  title   = {Lead-Free Halide Double Perovskite Materials: A New Superstar Toward Green and Stable Optoelectronic Applications},
  journal = {Nano-Micro Letters},
  year    = {2019},
  volume  = {11},
  pages   = {16},
  doi     = {10.1007/s40820-019-0244-6}
}

@article{Dengo2025,
  author  = {Nicola Dengo and David F. Macias-Pinilla and Pietro Anzini and Mara Colombo and Simone Virga and Andrea Brambilla and Piero Antonio Zecca and Damiano Monticelli and Francesco Giannici and Federica Bertolotti},
  title   = {Atomistic Insights into Halide Double Perovskite Nanocrystals Obtained by Multistep Synthesis and Efficient Compositional Engineering},
  journal = {ACS Nano},
  year    = {2025},
  volume  = {19},
  number  = {33},
  pages   = {30151--30164},
  doi     = {10.1021/acsnano.5c06497}
}

@article{Wang2022,
author = {Wang, Chao and Sun, Mengjiao and Wang, Hui and Zhao, Guangjiu},
doi = {10.1021/acs.jpclett.2c03482},
issn = {1948-7185},
journal = {The Journal of Physical Chemistry Letters},
month = {dec},
number = {1},
pages = {164--169},
title = {{Cubic Halide Double Perovskite Nanocrystals with Anisotropic Free Excitons and Self-Trapped Exciton Photoluminescence}},
url = {https://doi.org/10.1021/acs.jpclett.2c03482},
volume = {14},
year = {2022}
}

@article{locardi2019emissive,
  title={Emissive Bi-doped double perovskite Cs2Ag1--x Na x InCl6 nanocrystals},
  author={Locardi, Federico and Sartori, Emanuela and Buha, Joka and Zito, Juliette and Prato, Mirko and Pinchetti, Valerio and Zaffalon, Matteo L and Ferretti, Maurizio and Brovelli, Sergio and Infante, Ivan and others},
  journal={ACS Energy Letters},
  volume={4},
  number={8},
  pages={1976--1982},
  year={2019},
  publisher={ACS Publications}
}

@article{shaek2023does,
  title={How Does Local Strain Affect Stokes Shifts in Halide Double Perovskite Nanocrystals?},
  author={Shaek, Saar and Khalfin, Sasha and Massasa, Emma Hasina and Lang, Arad and Levy, Shai and Kortstee, Lotte TJ and Shamaev, Betty and Dror, Shaked and Lifer, Rachel and Shechter, Reut and others},
  journal={Chemistry of Materials},
  volume={35},
  number={21},
  pages={9064--9072},
  year={2023},
  publisher={ACS Publications}
}

@article{kepenekian2017rashba,
  title={Rashba and Dresselhaus couplings in halide perovskites: Accomplishments and opportunities for spintronics and spin--orbitronics},
  author={Kepenekian, Mika{\"e}l and Even, Jacky},
  journal={The Journal of Physical Chemistry Letters},
  volume={8},
  number={14},
  pages={3362--3370},
  year={2017},
  publisher={ACS Publications}
}

@article{baranowski2020excitons,
  title={Excitons in metal-halide perovskites},
  author={Baranowski, Michal and Plochocka, Paulina},
  journal={Advanced Energy Materials},
  volume={10},
  number={26},
  pages={1903659},
  year={2020},
  publisher={Wiley Online Library}
}

@article{quarti2024exciton,
  title={Exciton ground state fine structure and excited states landscape in layered halide perovskites from combined BSE simulations and symmetry analysis},
  author={Quarti, Claudio and Giorgi, Giacomo and Katan, Claudine and Even, Jacky and Palummo, Maurizia},
  journal={Advanced Optical Materials},
  volume={12},
  number={8},
  pages={2202801},
  year={2024},
  publisher={Wiley Online Library}
}

@article{filip2021phonon,
  title={Phonon screening of excitons in semiconductors: Halide perovskites and beyond},
  author={Filip, Marina R and Haber, Jonah B and Neaton, Jeffrey B},
  journal={Physical review letters},
  volume={127},
  number={6},
  pages={067401},
  year={2021},
  publisher={APS}
}

@article{baek2025exciton,
  title={Exciton Dynamics in Layered Halide Perovskite Light-Emitting Diodes},
  author={Baek, Sung-Doo and Yang, Seok Joo and Yang, Hanjun and Shao, Wenhao and Yang, Yu-Ting and Dou, Letian},
  journal={Advanced Materials},
  volume={37},
  number={25},
  pages={2411998},
  year={2025},
  publisher={Wiley Online Library}
}

@article{vegard1921konstitution,
  title={Die konstitution der mischkristalle und die raumf{\"u}llung der atome},
  author={Vegard, Lars},
  journal={Zeitschrift f{\"u}r Physik},
  volume={5},
  number={1},
  pages={17--26},
  year={1921},
  publisher={Springer}
}

@article{weber1953rotational,
  author  = {Weber, Gregorio},
  title   = {Rotational Brownian Motion and Polarization of the Fluorescence of Solutions},
  journal = {Advances in Protein Chemistry},
  volume  = {8},
  pages   = {415--459},
  year    = {1953},
  doi     = {10.1016/S0065-3233(08)60096-0}
}

@article{ameloot2013fluorescence,
  author  = {Ameloot, Marcel and vandeVen, Marleen and Acu{\~n}a, A. Ulises and Valeur, Bernard},
  title   = {Fluorescence anisotropy measurements in solution: Methods and reference materials},
  journal = {Pure and Applied Chemistry},
  volume  = {85},
  number  = {3},
  pages   = {589--608},
  year    = {2013},
  doi     = {10.1351/PAC-REP-12-09-12}
}

@article{lakowicz1999luminescence,
  author  = {Lakowicz, Joseph R. and Gryczynski, Ignacy and Gryczynski, Zygmunt and Murphy, Catherine J.},
  title   = {Luminescence Spectral Properties of CdS Nanoparticles},
  journal = {The Journal of Physical Chemistry B},
  year    = {1999},
  volume  = {103},
  number  = {36},
  pages   = {7613--7620},
  doi     = {10.1021/jp991469n},
  pmid    = {32267907},
  pmcid   = {PMC7106973}
}

@article{petrov2005timeresolved,
  author  = {Petrov, Eugene P. and Cichos, Frank and Zenkevich, Eduard and Starukhin, Dmitri and Shulga, Alexander},
  title   = {Time Resolved Photoluminescence Anisotropy of CdSe/ZnS Nanoparticles in Toluene at 300 K},
  journal = {Chemical Physics Letters},
  year    = {2005},
  volume  = {402},
  pages   = {233--238},
  publisher = {Elsevier},
  doi     = {10.1016/j.cplett.2004.12.019}
}

@article{Bekenstein2018,
  author  = {Bekenstein, Yehonadav and Dahl, Jakob C. and Huang, Jun and Osowiecki, Wojciech T. and Swabeck, Joseph K. and Chan, Emory M. and Yang, Peidong and Alivisatos, A. Paul},
  title   = {The Making and Breaking of Lead-Free Double Perovskite Nanocrystals of Cesium Silver--Bismuth Halide Compositions},
  journal = {Nano Letters},
  year    = {2018},
  volume  = {18},
  number  = {6},
  pages   = {3502--3508},
  doi     = {10.1021/acs.nanolett.8b00560}
}

@article{Khalfin2019,
  author  = {Khalfin, Sasha and Bekenstein, Yehonadav},
  title   = {Advances in Lead-Free Double Perovskite Nanocrystals, Engineering Band-Gaps and Enhancing Stability through Composition Tunability},
  journal = {Nanoscale},
  year    = {2019},
  volume  = {11},
  number  = {18},
  pages   = {8665--8679},
  doi     = {10.1039/C9NR01031A}
}

@misc{CrystalMaker,
  author    = {{CrystalMaker Software Ltd.}},
  title     = {CrystalMaker},
  version   = {11},
  address   = {Oxford, England},
  publisher = {CrystalMaker Software Ltd.},
  url       = {https://www.crystalmaker.com}
}

@misc{MaterialsProject_mp989571,
  author       = {{Materials Project}},
  title        = {Materials Data on Cs2NaInCl6 by Materials Project},
  howpublished = {Materials Project, mp-989571},
  note         = {Crystal structure data},
  year         = {2026}
}

\end{document}